\documentclass[12pt]{article}
\usepackage[dvips]{graphics}
\title{Physical stability of the QED vacuum}
\author{Hrvoje Nikoli\'c \\
Theoretical Physics Division, Rudjer Bo\v{s}kovi\'{c} Institute, \\
P.O.B. 180, HR-10002 Zagreb, Croatia \\
{\normalsize hrvoje@faust.irb.hr} \\
\makebox[1in]{} \\
}
\date{\today}
\begin{document}
\maketitle
\begin{abstract}
The possibility of electron-positron pair creation 
by an electric field is studied by using only those methods in field theory 
the predictions of which are confirmed experimentally. These 
methods include the perturbative method of quantum electrodynamics  
and the bases of classical electrodynamics. Such an approach includes 
the back reaction. It is found that the 
vacuum is always stable, in the sense that pair creation, 
if occurs, cannot be interpreted as a decay of the vacuum, but rather 
as a decay of the source of the electric field or as a 
process similar to bremsstrahlung. It is also found that 
there is no pair creation in a static
electric field, because it is inconsistent with energy conservation.
The non-perturbative aspects arising from the Borel summation of a 
divergent perturbative expansion are discussed. 
It is argued that the conventional methods that predict 
pair creation in a classical background electric field 
cannot serve even as approximations. The analogy 
with the possibility of particle creation by a gravitational 
field is qualitatively discussed.     
\end{abstract}
\vspace*{0.5cm}


\section{Introduction}

From the phenomenological point of view, one of the most important 
utilities of quantum field theory is the description of  
creation and destruction of particles
in physical processes. On the other hand, from the 
theoretical point of view, the concept of particles is not a 
well-defined concept in quantum field theory. This is because 
the operator of the number of particles is not defined in 
terms of local quantum fields and their derivatives. For 
free fields, it is defined in terms of raising and lowering 
operators. This definition is not unique because it depends on 
the choice of the complete orthogonal set of solutions to the 
free equations of motion. Although there exists a 
``natural" choice of solutions for free fields, the resulting 
definition of particles cannot be generalized uniquely when 
an interaction is present. Therefore, 
from the theoretical point of view, {\em all methods that 
describe creation and destruction of particles in physical 
processes are more or less vague}. 
From the pragmatic point of view, the theoretical problems 
with a method are not important if this method gives results 
that agree with experimental observations. 

There exist several independent theoretical indications 
that strong electric fields cause electron-positron pair creation. 
However, this 
effect has never been measured and {\em the methods used to 
describe this effect have never been justified experimentally}.  
In \cite{schw}, the imaginary part of the effective action of 
a quantum Dirac field interacting with a fixed classical 
electric field is found to be larger than zero,  
indicating that the absolute value of the vacuum-to-vacuum 
transition amplitude is smaller than 1. Although this 
gauge-independent description of the vacuum instability is 
often viewed as the clearest argument that a constant electric 
field causes pair creation, it has never been confirmed 
experimentally that the method used in \cite{schw} 
to calculate the 
{\em imaginary} part of the effective action correctly describes
any physical effect. Pair creation by an electric field 
is also predicted by the Bogoliubov-transformation 
method \cite{man,padmprl}. However, no physical effect  
resulting from a nontrivial Bogoliubov transformation has ever 
been observed. A tunneling picture can also describe 
vacuum instabilities, including pair creation by an electric field 
\cite{padm}. Again, the tunneling picture used to describe 
particle creation and destruction has never been justified 
experimentally. (Of course, it is experimentally confirmed that 
the tunneling picture may be applied to certain 
nonrelativistic quantum effects in which 
the number of particles does not change.)

Actually, the only experimentally justified 
method in quantum field theory that describes 
creation and destruction of particles is the perturbative 
method described by Feynman diagrams. Therefore, the most 
reliable approach to the study of the possibility of  
pair creation by an electric field seems to be this method. 
In Sec.~2 we use Feynman diagrams to study the possibility 
of pair creation in the field of a pointlike source. 
The possibility of pair creation by an electric field of a 
macroscopic source is discussed in Sec.~3. In Sec.~4 we 
prove that the QED vacuum is stable in a static 
classical background electric 
field and argue that the methods that 
predict an unstable vacuum in a classical background electric 
field cannot serve even as approximations. In Sec.~5 we 
discuss
the non-perturbative aspects arising from the Borel summation of the 
divergent perturbative expansion and criticize the usual 
formal procedure that predicts an unstable vacuum by  
transforming an infinite real quantity into a 
finite complex quantity. The analogy 
with the possibility of particle creation by gravitational 
fields is qualitatively discussed in Sec.~6, while the 
conclusions are drawn in Sec.~7.
   
\section{The description in terms of Feynman diagrams}

Our basic philosophy is to use only those methods 
the predictions of which are confirmed experimentally. 
Pair creation by an electric field, as well as any 
kind of vacuum instabilities, has never been observed. 
Therefore, in this section we base our analysis on an 
analogy with the observed effects the theoretical description 
of which is similar to the corresponding description 
of pair creation. 

Let us start with the discussion of elastic electron scattering 
in a classical background electric field. 
At the first order of perturbation theory, it is described 
by the diagram in Fig.~1(a). We assume that the electric 
field is static and vanishes at infinity, which implies that 
the initial energy of the electron at infinity is equal 
to the final energy of the electron at infinity. However, 
since the presence of the electric field breaks the translational 
symmetry, the initial 3-momentum ${\bf p}_{\rm i}$ does not need to be 
equal to the final 3-momentum ${\bf p}_{\rm f}$. 

Of course, the 3-momentum of the whole system must be conserved. 
The picture in which the electric field is a fixed classical 
bacground is only an approximation. First, the change of 
the electron 3-momentum causes a back reaction on the 
electric field. Second, the electromagnetic field  
should also be quantized. As far as the initial and final asymptotic 
states are all that 
interests us, a consistent description with a quantized  
electromagnetic field and the back reaction included is described 
by the diagram in Fig.~1(b). It is manifest in this diagram 
that the 3-momentum is conserved because the charged particle 
represented by the double line recoils. A fixed 
classical electromagnetic field 
is now replaced by a ``virtual photon", i.e., by the propagator 
of the electromagnetic field
\begin{equation}\label{prop}
D_{\mu\nu}(q)=-\frac{1}{q^2}\left( g_{\mu\nu}+(a-1)
\frac{q_{\mu}q_{\nu}}{q^2} \right) \; .
\end{equation}
The parameter $a$ parametrizes the class of gauges that satisfy 
the Lorentz condition $\partial_{\mu}A^{\mu}=0$. Although 
we do not know a general explicit proof that the physical effects of 
the electromagnetic-field quantization do not depend on the 
choice of gauge, it is well known that these effects do not depend 
on the choice of $a$, i.e., on the choice of gauge that satisfies 
the Lorentz condition. Besides, {\em the phenomenological 
predictions of QED based on the propagator (\ref{prop}) 
agree with experimental observations.} 

If the mass $M$ of the particle represented by the double line in 
Fig.~1(b) is much larger than the electron mass $m$ 
(for example, this heavy particle may be a proton treated as an 
elementary particle or a muon), then a static  
approximation is reasonable. In this approximation, 
both the initial and the final 3-momentum 
of the heavy particle are equal to zero. Consequently, in the 
Feynman gauge $a=1$, the effect of the propagator 
(\ref{prop}) is the same as that of the fixed classical 
electromagnetic background described 
(in the momentum space) by the electromagnetic 
potential $A^i=0$, $A^0\propto {\bf q}^{-2}$. This corresponds 
to a static electric field caused by a pointlike charged source. 
In this way, the approximation corresponding to 
Fig.~1(a) is derived from an ``exact" result 
(at the first order of perturbation theory) 
represented by Fig.~1(b). Note that the calculation 
of the diagram in Fig.~1(a) depends on the choice of gauge 
that satisfies the Lorentz condition and describes a given 
electric field. Therefore, the ``exact" diagram in  
Fig.~1(b) tells us that the ``correct" gauge in 
Fig.~1(a) describing a static electric field is
\begin{equation}\label{gauge} 
A^{\mu}=(A^0({\bf x}),0,0,0) \; .
\end{equation}        

The physics of an electron scattering in the field of a 
pointlike massive charge can be understood even with 
classical physics. Since pair creation is a 
genuine quantum effect, it is instructive to discuss the  
$\gamma\rightarrow e^+ e^-$ process in the electric 
field of a nucleus, as an example of a {\em measured} \cite{wrig} 
genuine quantum effect, similar to 
the $e^+ e^-$ creation by an electric field. It is described by 
the diagram in Fig.~2(a). Assuming again that the 
electric field is static and vanishes at infinity, we see that 
the energy is conserved again, while the 3-momentum is not. 
The ``exact" diagram, in which the 3-momentum is also conserved, 
is shown in Fig.~2(b). Similarly to the case of electron 
scattering, we see that the final effect of the back reaction 
is a recoil of the heavy source of the electric field. 

We are now ready to discuss the possibility of  
pair creation by an electric field. It is described by 
the diagram in Fig.~3(a). From the kinematics, it is easy 
to see that such a pair creation may be consistent with 
the 3-momentum conservation, but cannot be consistent 
with the energy conservation. On the other hand, if 
the electric field is static and vanishes at 
infinity, then energy must be conserved. In other words, 
the pair cannot be created, simply because the 
kinematics forbids it. This fact is even clearer from 
the ``exact" diagram in Fig.~3(b). 

Actually, there is a way to make the diagram in Fig.~3(b) 
consistent with the 4-momentum conservation, 
provided that the final heavy mass $M_{\rm f}$ is not equal to the 
initial heavy mass $M_{\rm i}$. (The usual QED Lagrangian does not 
contain such a vertex, but an effective Lagrangian for a composite particle
such as a nucleus could contain it.)
Obviously, the process is 
possible if $M_{\rm i} \geq M_{\rm f}+2m$. In that sense, 
pair creation is possible. However, it does not mean that 
the vacuum may be unstable. Since the mass of the 
source of the ``classical background" changes during the 
process, this process is naturally interpreted as a 
spontaneous decay of a particle with the mass $M_{\rm i}$ into 
the $e^+ e^-$ pair and a particle with the mass $M_{\rm f}$. 
If the initial heavy particle is stable, then the electric 
field of this particle does not create $e^+ e^-$ pairs.

Another way of making the diagram in Fig.~3(b) consistent 
with the 4-momentum conservation is to assume that 
the double-line
particle is accelerated by some external force. 
The process is similar to bremsstrahlung, in which
outgoing photons are replaced by outgoing electron-positron
pairs. In this case,
the energy needed for pair creation comes from the agency that
accelerates the double-line particle.
  
\section{Pair creation by a macroscopic electric field}

In the preceding section we have seen that 
the back reaction is automatically included when the 
electromagnetic field and its source are quantized. We 
have also seen that all effects of this back reaction 
are easily understood by classical kinematics related to 
the 4-momentum conservation.
On the other hand, when the source of the 
electromagnetic field is not one particle, but a huge 
number of particles, then it is not easy to describe 
the electromagnetic field and its source by a 
quantum formalism. However, the interactions among particles 
are essentially the same, so, again, one can apply the classical 
conservation laws in order to determine whether pair creation 
is possible. The only significant difference is the 
fact that particles that constitute the macroscopic 
source may not be separated in the final and initial state, so 
one cannot say that the initial and final 
states of particles are free states. Consequently,  
not all consequences of perturbative techniques can be applied.  
However, the basic understanding of classical electrodynamics is 
sufficient to give a qualitative description of the macroscopic 
electromagnetic field and its source.  

Consider, for example, 
the possibility of pair creation by 
an electric field produced by a  capacitor  
consisting of two large, parallel, and oppositely charged 
conducting plates. 
Such a configuration may be relevant to an experimental 
investigation of pair creation. Since the charge of the 
plates is fixed,  
the electric field between the plates depends only on the 
relative position of the plates. Since  
this field may be approximated by a constant electric field 
between the plates, one could expect that the pairs are 
created \cite{schw,man,padmprl,padm} with a thermal 
distribution in energy \cite{step,par1,par2}. However, 
the pair creation cannot be consistent with the conservation 
of energy. If the pair is detected far away from the 
capacitor, then the presence of the pair does not influence the energy 
of the plates and the energy of the electric field between them. 
Therefore, the back reaction cannot save the conservation of 
energy, implying that the pair is not created. 

By changing some experimental conditions, there is still a 
possibility for pair creation without a decay of  
particles that constitute the plates. Since the pair 
creation increases the energy of the system, the creation should 
be accompained by an decrease of the capacitor energy. 
The capacitor energy will decrease if the plates of the 
capacitor come closer. However, if the position 
of the plates is fixed (by some strong mechanical force), 
then the plates cannot come closer and there will be no 
pair creation. On the other hand, if the 
plates are not fixed, then the electric field will attract 
them, converting the energy of the electric field into the 
kinetic energy of the plates. If pair creation also 
occurs during this collapse of the plates, then the conservation 
of energy implies that the increase of the kinetic energy 
of the plates will be smaller than the decrease of the field 
energy. Therefore, on the macroscopic level, 
{\em pair creation is a dissipative process 
that manifests as a friction force} that acts on the plates.
This supports the similarity with bremsstrahlung, 
which is also a dissipative process that acts as a friction. 

The pair creation by a capacitor with unfixed plates can also  
be interpreted as a decay. In this interpretation, it is the 
capacitor itself that decays into a lower-energy state, because 
the initial state has a large 
potential energy stored in the electric field, while the final state, 
in which the plates are close to each other, has much lower potential 
energy. Part of this energy is transmitted into the produced pairs.               

\section{Vacuum-to-vacuum transition amplitude}

In the preceding sections we have seen that pair creation in an 
electric field is possible 
under certain circumstances. If the source of the field is a 
pointlike particle, then the probability and distribution of 
the created pairs can be calculated by well-understood 
and experimentally justified methods described by Feynman diagrams. 
Quantum electrodynamics {\em without} an external classical 
electromagnetic background is sufficient, but the description based on  
a classical background may serve as an approximation. 

If the source of the electric field consists of a huge number of particles, 
then the calculation of Feynman diagrams is not an efficient 
method for calculating the effect. Our discussion does not suggest 
how to calculate it in this case. One could guess that the 
conventional methods \cite{schw}-\cite{padm} that describe 
pair creation by classical electric fields are good 
approximations. Although they are not consistent with the 
conservation of energy, one could assume that these methods may 
serve as an adiabatic approximation, because the pair creation is 
slow, so the violation of the energy-conservation law is small. 
Assuming this, one can modify the method by introducing the 
effects of back reaction \cite{coop}-\cite{klug2}. In this section,  
we argue that the methods \cite{schw}-\cite{padm} are 
completely wrong, i.e., they cannot serve even as approximations. 

Take, for example, an approximation in which the Dirac field is 
quantized and interacts with a fixed static electric field. 
The method presented in \cite{schw} predicts that the absolute 
value of the vacuum-to-vacuum transition amplitude is smaller 
than one, suggesting an unstable vacuum. On the other hand, 
the method based on Feynman diagrams shows that the vacuum 
cannot decay simply because the amplitudes 
$\langle n,{\rm out}|0,{\rm in}\rangle \equiv 
\langle n|S|0 \rangle$ vanish for all physical on-shell states 
$|n\rangle$ orthogonal to the vacuum $|0\rangle$. The unitarity 
implies that the Feynman-diagrams approach must give 
\begin{equation}\label{stab}
|\langle 0|S|0 \rangle |=1 \; ,
\end{equation}
in contradiction with the result of \cite{schw}. Let us show 
explicitly that (\ref{stab}) is the correct result.

The Lagrangian of the system is 
\begin{equation}
{\cal L}={\cal L}_{\rm free}+{\cal L}_{\rm int} \; ,
\end{equation}
where 
\begin{eqnarray}
& {\cal L}_{\rm free}=
 \bar{\psi}(i\gamma^{\mu}\partial_{\mu}-m)\psi \; , & 
 \nonumber \\
& {\cal L}_{\rm int}=-e\bar{\psi}\gamma_{\mu}\psi A^{\mu} \; . &
\end{eqnarray}
The background $A^{\mu}$ is of the form (\ref{gauge}).
Since we use the interaction picture,    
the quantum Dirac field satisfies the free Dirac equation, 
so it can be expanded as 
\begin{eqnarray}\label{exp}
\psi(x) & = & \displaystyle\sum_{s}\int\frac{d^3k}{(2\pi)^{3/2}}
\sqrt{\frac{m}{\omega(k)}} \nonumber \\
& & \times [ b_s(k)u_s(k)e^{-ikx}+
d^{\dagger}_s(k)v_s(k)e^{ikx}] ,
\end{eqnarray}
where $u_s(k)$ and $v_s(k)$ are free spinors \cite{bjorkdr}, 
$b_s(k)$ destroys electrons, and $d^{\dagger}_s(k)$ creates 
positrons. The S-matrix is given by 
\begin{equation}\label{S}
S=Te^{i\int d^4x {\cal L}_{\rm int}} \; .
\end{equation}
Therefore, the exponent in (\ref{S}) is proportional to
\begin{equation}\label{int} 
\int d^3x \, A^{0}({\bf x}) \int_{-\infty}^{\infty} dt\,  
\bar{\psi}({\bf x},t)\gamma_0 \psi({\bf x},t) \; .
\end{equation}
When the expansion (\ref{exp}) is used in (\ref{int}), then 
four types of terms appear, i.e., the terms proportional to 
$b^{\dagger}_s(k)b_{s'}(k')$, $d_s(k)d^{\dagger}_{s'}(k')$, 
$b^{\dagger}_s(k)d^{\dagger}_{s'}(k')$, or 
$d_s(k)b_{s'}(k')$. (The time-ordering changes the ordering of the 
operators, which we ignore because it does not influence our 
conclusions.)  
This implies that $S|0\rangle$ is a 
fermion variant \cite{chat,fan} of a  
squeezed state \cite{schum,gris}. The second two types 
of terms are responsible for the squeezing, i.e., for 
the nontrivial particle content of the state $S|0\rangle$.
If these terms are missing, then the  
first two types of terms only change the phase of the 
vacuum. 

The time integration can be performed 
before the 3-momentum integrations. Therefore, the 
time integration of the first two types of terms leads to a 
factor proportional to $\delta(\omega-\omega')$, while that 
of the second two types of terms leads to a 
factor proportional to $\delta(\omega+\omega')$. Since 
$\omega+\omega'$ cannot be zero, the integration over the 
3-momenta kills the second two types of terms. Therefore, 
only the first two types of terms appear in the final 
expression, which implies that  
the $S$-matrix operator does not change the number of particles 
when acts on a state with a definite number of particles. 
The function $\delta(\omega-\omega')$ provides that the 
energy is conserved in any process described by the 
$S$-matrix element $\langle {\rm f}|S|{\rm i}\rangle$. 
When $S$ acts on the vacuum, it merely changes its phase. 
This proves Eq. (\ref{stab}), i.e., the stability of the 
vacuum in a static electric field. It also clearly 
shows that this stability is directly related to the 
conservation of energy, i.e., to the factor 
$\delta(\omega+\omega')$ which appears because $A^{\mu}$ 
does not depend on time.  

Recall that the method used above to prove (\ref{stab}) is 
justified by experiments, while there is no such  
justification for the method used in \cite{schw} 
to show that (\ref{stab}) is not true. It is also 
important to stress that, although the method in 
\cite{schw} is very different from the method used 
above, the {\em physical} assumptions and approximations 
are the same. Therefore, one of the {\em methods} must be 
completely wrong; different results cannot be 
manifestations of different physical assumptions and approximations. 
All remarks made in this paper  
clearly indicate that it is the method in \cite{schw} that 
must be wrong. Actually, it has already been shown by an independent 
argument that the method presented in \cite{schw} is   
inconsistent \cite{nik2}.     
  
Our discussion suggests that other methods \cite{man,padmprl,padm} 
that predict pair creation by a classical 
static background electric field should also be wrong. 
The inconsistencies of these methods have  
also been discussed by independent arguments \cite{nik2,nik1}.

\section{Non-perturbative aspects}

One could argue that the result of Sec. 4 differs from that of 
\cite{schw} (for a constant electric field)
because the latter is a non-perturbative result, 
while the former is based on the perturbative expansion. 
(By a perturbative result we mean a mathematical expression 
written as a series expansion in $e$ without negative powers of $e$.)  
In other words, one could object that the calculation in Sec. 4 
is wrong because the non-perturbative contributions are not 
included in this calculation. However, as shown in \cite{dunne}, 
{\em the non-perturbative result of \cite{schw} does not arise 
from a contribution which is not present in a perturbative expansion}.
The calculation in \cite{dunne} is not based on the 
interaction picture, but, as we demonstrate below, essentially 
the same arguments can be applied to the calculation based on the 
interaction picture of Sec. 4.  

By calculating all contributions to $\langle 0| S|0\rangle$ for 
$A^{\mu}$ of the form (\ref{gauge}), we obtain
\begin{equation}\label{perturb}
\langle 0| S|0\rangle =\exp [iW(e)] \; ,
\end{equation}
where $W(e)$ is given by the sum of all different one-loop 
diagrams withot external legs. It can be written as
\begin{equation}\label{perturb2}
W(e)=\sum_{n=0}^{\infty} c_n e^n \; ,
\end{equation}
where $c_n$ are functionals of $A^{0}({\bf x})$. They are real because 
$A^{0}({\bf x})$ is real. To understand how a non-perturbative result may 
arise from a perturbative result (\ref{perturb2}), it is not necessary 
to calculate $c_n$ explicitly. It is crucial to know that 
the series (\ref{perturb2}) is divergent, even if the coefficients 
$c_n$ are made finite by some regularization or renormalization 
procedure. Assume, for example, that they are of the form \cite{dunne}
\begin{equation}\label{perturb3}
c_n = \alpha^n n! \; ,
\end{equation}
where $\alpha$ is some real positive constant. Clearly, the series 
(\ref{perturb2}) with coefficients (\ref{perturb3}) has a zero 
radius of convergence. Since the coefficients do not alternate 
in sign, even the Borel summation does not give a convergent result. 
Nevertheless, let us do the formal Borel summation. By writing 
\begin{equation}\label{faktorijela}
n! =\int_{0}^{\infty} ds \, s^n \exp(-s) \; ,
\end{equation}
putting this in (\ref{perturb3}), and interchanging the order of 
summation and integration in (\ref{perturb2}), we obtain 
\begin{equation}\label{noperturb}
W(e)=\frac{1}{\alpha e}\int_{0}^{\infty} dz \, \frac{1}{1-z}
\exp \left( -\frac{z}{\alpha e} \right) \; ,
\end{equation}
where a new integration variable is introduced:
\begin{equation}\label{var}
z=\alpha e s \; . 
\end{equation}
If (\ref{noperturb}) is expanded in powers of $e$
and (\ref{var}) is ignored, then negative 
powers of $e$ appear. In other words, (\ref{noperturb}) is a 
{\em non-perturbative} result. We also see that the formal 
Borel summation given by (\ref{noperturb}) 
did not change the fact that $W(e)$ is {\em real}, 
corresponding to a stable vacuum in (\ref{perturb}). 

One could be worried by the fact that (\ref{noperturb}) is 
{\em divergent}, which is a consequence of the pole of the 
subintegral function at $z=1$. Therefore, one could avoid the 
pole by deforming the contour of integration \cite{dunne}, 
which leads to a finite real part of $W(e)$ and a 
positive {\em imaginary} part of $W(e)$. If such a formal 
procedure of giving an imaginary part to a real 
quantity makes sence, then it implies that (\ref{noperturb}) 
corresponds to an {\em unstable} vacuum. However, it has  
already been shown that such an artificial formal procedure is not 
consistent \cite{nik2}. Below we give new physical arguments 
against such a deformation of the integration contour.     
 
First, the fact that $W(e)$ is divergent is not a problem. 
The physical quantity calculated from (\ref{perturb}) is 
$|\langle 0| S|0\rangle|^2$. It is equal to 1 for any real 
number $W$, including the case in which $W$ approaches 
infinity along the real axis. Therefore, there is no a 
physical reason to deform the integration contour. 
Moreover, the fact that (\ref{noperturb}) is divergent does 
not imply that it is infinite. Actually, the infinite 
contribution from $z=1-\epsilon$ (with $\epsilon\rightarrow 0^+$) 
is canceled by the infinite 
contribution from $z=1+\epsilon$. In other words, the principal 
value of the integral in (\ref{noperturb}) is {\em real} and 
{\em finite}. The same is true for similar divergent integrals that 
appear in \cite{schw} and \cite{dunne} related to 
the explicit calculation of the vacuum-to-vacuum transition 
amplitude in a constant background electric field. This 
finite and real principal value is the most natural 
choice for these divergent integrals, which leads to a 
stable vacuum.  

Second, assume that the contour should be deformed, so that the vacuum 
is unstable. As we have seen, although (\ref{noperturb}) is a 
non-perturbative result, all contributions to it arise from 
the Feynman diagrams. If the vacuum instability realizes as a 
pair creation, then there must be a non-zero probability 
of producing any particular number of pairs. For example,  
the probability 
of producing one pair is given by the diagram in Fig. 3(a).
However, there is no a consistent way of making this  
diagram non-vanishing for $A^{\mu}$ of the form (\ref{gauge}). 
In particular, this would be in contradiction with the 
conservation of energy. The ``need" for a non-zero 
value arising from the diagram in Fig. 3(a) does not emerge 
from any particular diagram contributing to (\ref{perturb2}), 
but from the whole diverging sum. Therefore, avoiding this 
divergence by an integration-contour deformation does not 
manifest as a modification of particular diagrams. 
In other words, it is not consistent to replace infinite real 
quantities by finite complex quantities in a way described above. 

To summarize, the non-perturbative aspects are artefacts of 
Borel summation of a divergent series, while the vacuum 
``instability" is an artefact of an integration-contour deformation.
The divergence of $W(e)$ does not cause any physical 
problems, while the integration-contour deformation is 
physically unjustified. Therefore, in our opinion, 
the physical interpretation 
of formal results in \cite{schw} and \cite{dunne} related to 
vacuum instability in a constant background 
electric field is wrong. 
   
\section{Analogy with the gravitational particle creation}

There is a lot of similarity between the particle creation 
by a classical electromagnetic background and that by 
a classical gravitational background 
\cite{padmprl,padm,step,par2,brout}. Unfortunately, it is 
still not known how to quantize gravity consistently, so 
it is not clear if it would be consistent to base a  
quantitative analysis on a diagram analogous to the diagram in 
Fig.~3(b). Nevertheless, by analogy, we can use the final 
results of the preceding sections to draw some qualitative 
conclusions related to the gravitational particle creation.  
   
The particle creation by a capacitor has many similarities 
to that by a black hole. First, since the electric field 
between the plates of the capacitor is approximately constant, 
one could expect a thermal distribution of the particles 
created \cite{step,par1,par2}, just as is expected 
for black holes \cite{hawk,bd}. However, we have shown that a 
static capacitor does not create particles; only a collapsing 
capacitor can do it. Similarly, we expect that if some 
mechanism prevents the gravitational collapse, then the black 
hole does not create particles.  

In analogy with the electromagnetic case, we may also speculate
about the effect of the back reaction related to the particle
creation by the source of the gravitational field. Our discussion
suggests that the back reaction manifests as a friction on the
moving matter, which slows down the gravitational collapse.
If a final state of matter (determined by the yet
unknown physics on the Planck scale) is reached, then the collapse
and the particle creation stops.

Another interesting question related to particle creation by 
gravitational fields, especially by black holes, is {\em where} 
particles are created. The analogy with the 
electromagnetic case suggests that the matter, i.e., the source of the 
gravitational field, is the source of the created particles.
If this is true, then {\em the created particles that escape from the 
black hole are created by matter that has not yet fallen into the  
hole bounded by the horizon}. This picture is drastically 
different from the usual picture that describes the 
black-hole evaporation \cite{hawk,bd}. Nevertheless, although the 
mechanism of particle creation by black holes 
suggested by our discussion is completely different from the usual one, we 
still expect a thermal distribution of escaped particles, as seen by a 
distant observer. This is because the thermal distribution is a 
consequence of the exponential red shift and can be understood 
even by classical physics \cite{pad3,pad4}, without any assumption 
on the physical mechanism that causes particle creation near the 
horizon.      

If the analogy with bremsstrahlung is used, then, since the 
matter moves inertially in the gravitational field, particle 
creation may not exist at all. Such a reasoning is reasonable if 
the equivalence principle is applicable on the level of quantum 
gravity. 

As we see now, the predictions of the conventional methods that 
describe particle creation by classical gravitational 
backgrounds may be completely wrong. The inconsistencies of these  
methods have already been discussed by independent arguments 
\cite{nik2,nik1,nik3}. A reliable prediction can be given only by 
a quantum-gravity theory that automatically includes the effects of 
back reaction.   

\section{Conclusions}

Pair creation by an electromagnetic field can be consistently 
studied only if the electromagnetic field and its source are 
quantized. Using the experimentally verified method of 
Feynman diagrams, we have found that the pairs are not created 
if the source of the field is a stable unaccelerating particle. 
Similarly, if the source of the field is a macroscopic static 
object, then the pairs are not created either. In certain cases  
there is the possibility of pair creation, but such an effect 
should not be interpreted as a vacuum instability. Instead, 
such a process can be interpreted as a decay of the source 
of the electromagnetic field or as a process similar to the 
bremsstrahlung. Although these 
results may seem to be trivial consequences of 
perturbative QED and the 
energy-conservation law, these results clearly indicate that 
the methods that predict pair creation by classical background 
electromagnetic fields are inappropriate. They cannot serve 
even as approximations. 

Analogous remarks are also valid for the particle creation by 
a gravitational field. In particular, this implies that 
our understanding of black-hole thermodynamics should be 
drastically revised.      

Yet, an approximation in which quantum particles move in classical 
electromagnetic and gravitational backgrounds should certainly make 
sense in some cases. A consistent general-covariant and gauge-invariant 
approximation that describes the particle content of quantum fields 
in classical backgrounds will be given elsewhere. 

\section*{Acknowledgement}
This work was supported by the Ministry of Science and Technology of the
Republic of Croatia under Contract No. 00980102.

\newpage


\begin{figure}
\centerline{\includegraphics{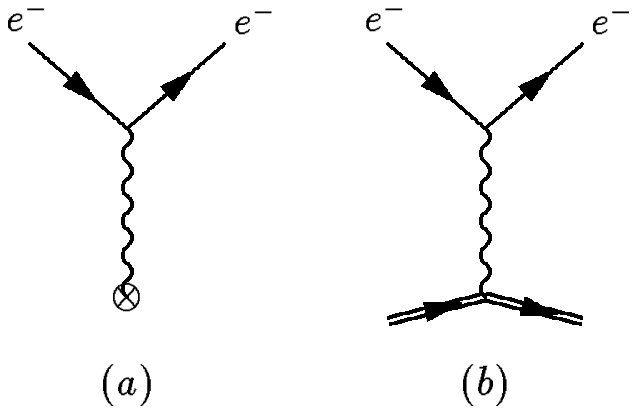}}
\caption{The lowest-order diagrams representing
elastic electron scattering caused by
the electromagnetic interaction. In diagram (a), the
electromagnetic interaction is described by a fixed classical
electromagnetic background. In diagram (b), the
electromagnetic interaction is described by a quantum virtual
photon. When the double-line particle is much heavier than
the electron, then diagram (b) can be approximated by
diagram (a) with a potential $A^{\mu}=(A^0({\bf x}),0,0,0)$ describing
the electric field of a pointlike charge.}
\end{figure}

\begin{figure}
\centerline{\includegraphics{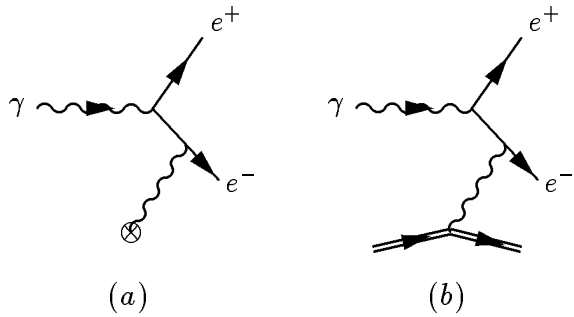}}
\caption{The lowest-order diagrams representing the 
$\gamma\rightarrow e^+ e^-$ process in the field of a pointlike
charge. (The diagrams in which the wiggly line
without an arrow  is connected
with the upper fermion line are omitted.) As in Fig.~1,
diagram (a) represents an approximation of diagram (b).}
\end{figure}

\begin{figure}[h]
\centerline{\includegraphics{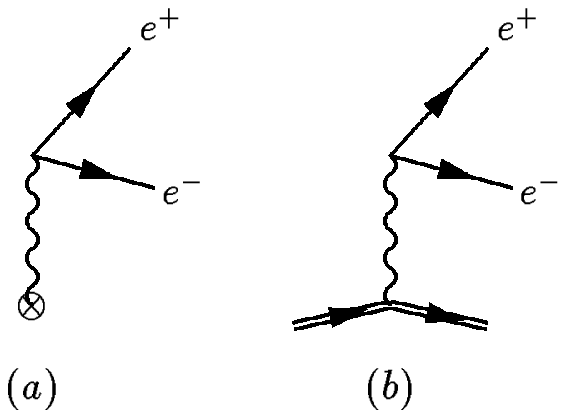}}
\caption{The lowest-order diagrams representing 
$e^+ e^-$ pair creation caused by the 
electromagnetic interaction. As in Fig.~1,
diagram (a) represents an approximation of diagram (b).
If the double line represents a particle for which the
final mass is equal to the initial mass, and if an external
force does not act on this particle, then,
owing to energy conservation, the amplitude
corresponding to diagram (b) vanishes.}
\end{figure}


\begin{thebibliography}{29}
\bibitem{schw}
J. Schwinger, Phys. Rev. 82 (1951) 664.
\bibitem{man}
C.A. Manogue, Ann. Phys. 181 (1988) 261.
\bibitem{padmprl}
T. Padmanabhan, Phys. Rev. Lett. 64 (1990) 2471.
\bibitem{padm}
K. Srinivasan, T. Padmanabhan, Phys. Rev. D 60 
(1999) 024007.
\bibitem{wrig}
L.E. Wright, K.K. Sud, D.W. Kosik, Phys. Rev. C 36 (1987) 562.
\bibitem{step}
C. Stephens, Ann. Phys. 193 (1989) 255. 
\bibitem{par1}
R. Brout, R. Parentani, Ph. Spindel, Nucl. Phys. B {\bf 353} 
(1991) 209.
\bibitem{par2}
R. Parentani, R. Brout, Nucl. Phys. B 388 (1992) 474.
%
\bibitem{coop}
F. Cooper, E. Mottola, Phys. Rev. D 40 (1989) 456.
\bibitem{bial}
I. Bialynicki-Birula, P. G\'ornicki, J. Rafelski, 
Phys. Rev. D 44 (1991) 1825.
\bibitem{klug}
Y. Kluger, J.M. Eisenberg, B. Svetitsky, F. Cooper,  
E. Mottola, Phys. Rev. D 45 (1992) 4659.
\bibitem{best}
C. Best, J.M. Eisenberg, Phys. Rev. D 47 (1993) 4639.
\bibitem{klug2}
Y. Kluger, E. Mottola, J.M. Eisenberg, Phys. Rev. D 58 
(1998) 125015.  
%
\bibitem{bjorkdr}
J.D. Bjorken, S.D. Drell, Relativistic Quantum Fields,  
McGraw-Hill, New York, 1965.
\bibitem{chat}
S. Chaturvedi, R. Sandhya, V. Srinivasan, Phys. Rev. A 41 
(1990) 3969.
\bibitem{fan}
H.-y. Fan, Y. Fan, F.T. Chan, Phys. Lett. A 247 (1998) 267.  
\bibitem{schum}
B.L. Schumacher, Phys. Rep. 135 (1986) 317.
\bibitem{gris}
L.P. Grishchuk, Y.V. Sidorov, Phys. Rev. D 42 (1990) 3413.
\bibitem{nik2}
H. Nikoli\'c, hep-th/0103251.
\bibitem{nik1}
H. Nikoli\'c, hep-th/0103053.
\bibitem{dunne}
G.V. Dunne, T.M. Hall, Phys. Rev. D 60 (1999) 065002.
\bibitem{brout}
R. Brout, S. Massar, R. Parentani, Ph. Spindel, 
Phys. Rep. 260 (1995) 329.
\bibitem{hawk}
S.W. Hawking, Commun. Math. Phys. 43 (1975) 199.
\bibitem{bd}
N.D. Birrell, P.C.W. Davies, Quantum fields in
curved space, Cambridge Press, NY, 1982.
\bibitem{pad3}
K. Srinivasan, T. Padmanabhan, gr-qc/9812087.
\bibitem{pad4}
M. Nouri-Zonoz, T. Padmanabhan, gr-qc/9812088.
\bibitem{nik3}
H. Nikoli\'c, Mod. Phys. Lett. A 16 (2001) 579; 
gr-qc/0103108.
\end{thebibliography}
\end{document}